\documentclass[a4paper,11pt]{article}
\pdfoutput=1 
\usepackage{aas_macros}
\usepackage{jcappub} 
\usepackage[T1]{fontenc} 
\usepackage{amsbsy}
\usepackage{color}
\usepackage{xcolor}
\usepackage{subcaption}
\usepackage{hyperref}
\usepackage{fontawesome}
\usepackage{multicol}

\usepackage[normalem]{ulem}

\newcommand{\be}{\begin{equation}}
\newcommand{\ee}{\end{equation}}

\usepackage{amssymb}

\title{\boldmath A primordial origin to cosmic tensions: towards reconciling $H_0$ and $S_8$ with early dark energy and scale-dependent primordial non-Gaussianities}

\author[a]{Cl\'ement Stahl,}
\author[b]{Vivian Poulin,}
\author[a]{Benoit Famaey,}
\author[a]{Rodrigo Ibata}

\affiliation[a]{Universit\'e de Strasbourg, CNRS, Observatoire astronomique de Strasbourg, UMR 7550, 67000 Strasbourg, France}

\affiliation[b]{Universit\'e de Montpellier, CNRS, Laboratoire Univers \& Particules de Montpellier (LUPM), UMR 5299, Montpellier, France}

\emailAdd{clement.stahl@unistra.fr}

\abstract{The Hubble ($H_0$) tension between direct measurements of the expansion rate and the prediction of the $\Lambda$CDM cosmological model calibrated on the Cosmic Microwave Background (CMB), is a strong motivation to explore alternative cosmological models. A popular class of such models includes an additional component of dark energy relevant in the early Universe, which solves the Hubble tension. These Early Dark Energy (EDE) models however typically overpredict the value of the $S_8$ parameter. Here, we show how combining EDE with scale-dependent primordial non-Gaussianities (sPNG) can in principle both increase $H_0$ and decrease $S_8$ at once, even conceivably allowing to solve the potential $S_8$ tension between measurements of weak gravitational lensing and the
$\Lambda$CDM expectation. Such sPNG are related to non-trivial inflationary physics, and observationally affect the non-linear regime of structure formation while leaving the linear regime mostly untouched. Depending on the amplitude of the sPNG, such models can either yield back the $\Lambda$CDM expectation for the power spectrum in the non-linear regime, and hence an $S_8$ parameter compatible with $\Lambda$CDM, or can even reconcile the value of $S_8$ from old weak-lensing measurements with the CMB, while solving the Hubble tension in all cases. In such models, both tensions would then be entirely related to pre-CMB physics of the early Universe.}

\begin{document}

\maketitle

\flushbottom

\section{Introduction}\label{sec:Introduction}

The $\Lambda$-Cold Dark Matter ($\Lambda$CDM) standard model of cosmology stems from the application of General Relativity to the Universe as a whole under the hypothesis of statistical homogeneity and isotropy. It relies on dark energy (in the form of a cosmological constant $\Lambda$) and Cold Dark Matter (CDM) generated at the end of the inflation era, forming an almost Gaussian distributed random field. The matter density contrast then underwent cosmic evolution to form the structures that we observe today. Despite its obvious successes, the model also faces a few tensions with observations. The main one is related to the background expansion --- the so-called Hubble ($H_0$) tension ---, whilst a less significant one is related to the amplitude of cosmological perturbations --- the so-called $S_8$ tension. 

The Hubble tension stems from the observation that the cosmic distance ladder --- built out of distance indicators such as supernovae of type 1a (SN1a) and baryonic acoustic oscillations (BAO) --- give different estimates for the current expansion rate of the universe $H_0$ when calibrated {\it directly} by model-independent inference of the SN1a intrinsic magnitude, or {\it indirectly} from model-dependent estimates of the sound horizon \cite{Verde:2019ivm,Abdalla:2022yfr,Verde:2023lmm}. The most significant tension ($>5\sigma$) between estimates of $H_0$ appears, on the one hand when using cepheid stars to calibrate SN1a as done by the S$H_0$ES collaboration \cite{Riess:2021jrx}, and on the other hand from the $\Lambda$CDM estimates of the sound horizon when fit to Planck Cosmic Microwave Background (CMB) data \cite{Planck:2018vyg}.
In this context, while systematic uncertainties are still actively looked for \cite{Freedman:2019jwv,Riess:2021jrx,Freedman:2024eph,Riess:2024vfa}, promising models to resolve this tension appear to be those that reduce the {\it physical} size of the sound horizon at recombination so as to compensate the effect of a larger $H_0$ on the angular diameter distance and leave the {\it angular} size of the sound horizon (that is very precisely measured) unaffected \cite{Schoneberg:2021qvd,Abdalla:2022yfr,Poulin:2024ken}. An example of such models is Early Dark Energy (EDE), in which a scalar field boosts the expansion rate from redshift $\sim 10^4$ to recombination. It hence reduces the sound horizon by increasing the value of the Hubble rate prior to recombination while keeping $H_0$ compatible with direct measurements \cite{Karwal:2016vyq,Poulin:2018cxd,Smith:2019ihp}. 
Yet, the dynamics of the EDE is highly constrained by current high-precision cosmological data (see Refs.~\cite{Kamionkowski:2022pkx,Poulin:2023lkg,McDonough:2023qcu} for recent reviews on the topic). Broadly speaking, the dynamics that appear in line with the data is such that the field is frozen in its potential by some mechanism (typically Hubble friction) and starts rolling down its potential around the redshift of matter radiation equality $z\sim 4000$. Models vary in the implementation of such a mechanism, with potential additional coupling to other species (e.g. Refs.~\cite{Sakstein:2019fmf,McDonough:2021pdg,Karwal:2021vpk,Liu:2023rvo,Liu:2023wew,Garcia-Arroyo:2024tqq,Simon:2024jmu}) or non-minimal coupling to gravity (e.g. Refs.~\cite{Braglia:2020auw,FrancoAbellan:2023gec}).

The $S_8$ parameter is defined as $S_8 = \sigma_8(\Omega_m/0.3)^{1/2}$, where $\Omega_m$ is the relative matter density at $z=0$ and $\sigma_8$ is the standard deviation of density fluctuations measured within spherical regions of 8~Mpc/h, encoding the amplitude of the power spectrum (PS). A tension was originally identified by the Kilo-Degree Survey (KiDS), which when combined with BOSS, yielded  $S_8 = 0.766_{-0.014}^{+0.020}$ \cite{Heymans:2020gsg} compared to the value expected in $\Lambda$CDM from Planck and SDSS BAO measurements  $S_8=0.825\pm0.011$ \cite{Planck:2018vyg}, with a discrepancy at the 3$\sigma$ level. Similarly, the Dark Energy Survey (DES) originally suggested a tension at the $2.4\sigma$ level, with $S_8=0.776\pm 0.017$.
Although subsequent reanalyses of KiDS, DES \cite{Kilo-DegreeSurvey:2023gfr} and Planck data \cite{2022arXiv220510869R,Tristram:2023haj} have yielded slightly different values for the $S_8$ parameter over the years, the tension between weak-lensing results and CMB-derived measurements remained present until recently. Barring potential systematic errors or subtle non-linear effects related to baryonic feedback\footnote{See also the small-scale weak-lensing data of the Hyper Suprime-Cam, that can be accounted for within dark matter only simulations \cite{Garcia-Garcia:2024gzy,Terasawa:2024agq}.} \citep[\textit{e.g.},][]{Amon:2022ycy,Amon:2022azi,Arico:2023ocu,Kilo-DegreeSurvey:2023gfr}, this potential $S_8$ tension may point to new dark matter (DM) properties, such as decay or interaction with baryons, with dark radiation or with dark energy \citep[see, \textit{e.g.},][]{Lesgourgues:2015wza,Buen-Abad:2015ova,Chacko:2016kgg,Buen-Abad:2017gxg,Heimersheim:2020aoc,DiValentino:2019ffd,Lucca:2021dxo,FrancoAbellan:2020xnr,DiValentino:2020vvd,Bansal:2021dfh,Baldi:2016zom,Kumar:2017bpv,Asghari:2019qld,BeltranJimenez:2020qdu,Figueruelo:2021elm,BeltranJimenez:2021wbq,Poulin:2022sgp,Simon:2024jmu}, or to some new non-linear effect potentially related to new degrees of freedom during inflation \cite{Stahl:2024stz}. The most recent reanalysis of KiDS data, with increased survey area and redshift depth however found $S_8 = 0.815_{-0.021}^{+0.016}$, hence with no tension \citep{Wright:2025xka,Stolzner:2025htz}.

Importantly, the success of EDE models to resolve the Hubble tension is hindered by the fact that they actually tend to increase the $S_8$ parameter. This is an inherent property of all models that resolve the Hubble tension by solely affecting the pre-recombination era and the value of the sound horizon, due to the fact that the cosmic distance ladder, when calibrated with the S$H_0$ES SN1a magnitude, provides a larger value of $\Omega_m h^2$ --- the physical density of matter --- than in $\Lambda$CDM. This leads to early matter domination and therefore a larger amplitude of fluctuations today \cite{Jedamzik:2020zmd,Blanchard:2022xkk,Poulin:2024ken,Pedrotti:2024kpn,Hill:2020osr,Ivanov:2020ril,DAmico:2020ods}, though the significance of the increase of $S_8$ is debated\footnote{It has been pointed out that the $S_8$ parameter may appear larger in EDE also because of the impact of a larger $h$ in the definition of $S_8$. The quantity $S_{\rm 12}$, where the scale $R=8h^{-1}$Mpc is replaced by an absolute scale $R=12$Mpc \cite{Sanchez:2020vvb}, appears in similar tension with EDE \cite{Secco:2022kqg,Forconi:2025cwp}.} \cite{Murgia:2020ryi,Smith:2020rxx,Herold:2021ksg,Forconi:2025cwp}. However, this relies on the assumption that the primordial power spectrum is described by a simple power-law, as in $\Lambda$CDM. Yet, there is a known important interplay between EDE and inflation, as the value of the primordial tilt $n_s$ favored in the EDE model is much closer to the scale-invariant value $n_s =1$ (at the $2\sigma$ level), potentially reshaping our understanding of cosmic inflation \cite{Takahashi:2021bti,Cruz:2022oqk}. Another aspect of all inflationary models, but with varying significance depending on the actual model, is the presence of Primordial Non-Gaussianities (PNG) \cite{Achucarro:2022qrl}. In the simplest realizations of inflation, with one (active) degree of freedom, PNG are scale invariant and very small \cite{Maldacena:2002vr}. But they can in principle be much larger and, in some exotic inflationary models \cite{LoVerde:2007ri, Byrnes:2009pe,Huang:2010es}, can be scale-dependent.

Observations from, e.g., the CMB \cite{Planck:2019kim} and Large Scale Structure (LSS) \cite{Chaussidon:2024zfp} have already placed tight constraints on PNG, which have to be small \textit{on large scales}. However, significant scale-dependent PNG (hereafter, sPNG) do have the potential to have an impact at smaller scales, relevant for the $S_8$ tension, while remaining consistent with these constraints. We have indeed pointed out \cite{Stahl:2024stz} that the original $S_8$ tension might be alleviated when considering sPNG leading to less structures in the non-linear regime at late times. In contrast with other potential solutions to this original $S_8$ tension, this non-linear modification of structure formation leaves the linear regime of cosmology untouched, indeed opening up the possibility to combine that solution with linear solutions to the Hubble tension such as those based on EDE. 

In this paper, we therefore simulate, for the first time, EDE and sPNG jointly in order to explore their potential to solve the $H_0$ and $S_8$ tensions at once. Combinations of EDE as a solution to the $H_0$ tension with other solutions to the $S_8$ tension have been studied elsewhere \cite{Clark:2021hlo,Simon:2024jmu}. Here, our choice illustrates that a well motivated deviation to the Gaussian primordial matter distribution can significantly alter strong bounds on EDE models coming from probes of the LSS \cite{Ivanov:2020ril,Hill:2020osr,Goldstein:2023gnw}. Therefore, the freedom there is in the form of the primordial matter distribution, ultimately connected with our currently limited constraints on the inflation era, should be seen as an important caveat to the robustness of LSS-driven constraints to EDE.

The paper is structured as follows.
In Section~\ref{sec:setups}, we review the salient features of EDE and sPNG models, and give details about our simulations setup. We then present in Section~\ref{sec:res} our results, and in particular how EDE+sPNG can in principle reconcile the $S_8$ value deduced at the level of the non-linear matter power spectrum with a high value of $H_0$, instead of inducing a tension in the case of EDE alone. We wrap up and give some perspectives in Section~\ref{sec:ccl}.

\section{Simulations of Early Dark Energy and Scale Dependent Primordial non-Gaussianities}\label{sec:setups}
\subsection{Early Dark Energy}
\label{sec:EDE}

In this paper, we will make use of the toy model of ``axion-like'' EDE which is defined by the potential 
\begin{equation}\label{eq:potential}
    V(\theta) = {\cal F}^2 m^2 [1-\cos (\theta)]^n,
\end{equation} 
where $m$ denotes the axion mass (${\cal O}\sim 10^{-28} eV$), ${\cal F}$ is the axion decay constant (${\cal O}\sim 0.1 m_{\rm pl}$), and $\theta$ is the dimensionless field variable constrained within $-\pi \leq \theta \leq \pi$. 
When the Hubble parameter $H$ drops to the scale of the field’s effective mass, $H \sim \partial^2_\theta V(\theta)$, the field begins oscillating around the minimum of its potential. The energy density associated with the field is thus constant until the field starts rolling (typically around matter-radiation equality), at which points it dilutes at a rate controlled by the exponent $n$, which we fix to $n=3$. The angular size of the sound horizon $\theta_s(z_*)\equiv r_s(z_*)/D_A(z_*)$, where $D_A$ is the angular diameter distance to recombination, is extremely precisely measured by CMB data \cite{Planck:2018vyg}. As it is proportional to $H_0$, a larger $H_0$ value would lead to a too large value compared to CMB measurements.
In our EDE model, the boost provided to the expansion history reduces the sound horizon at recombination $r_s(z_*)$, therefore allowing for a higher $H_0$ value. Note that it is common to perform analyzes on the related parameters $f_{\rm EDE}(z_c)$ representing the fractional contribution of the EDE component to the total energy density at the critical redshift $z_c$, below which the EDE dilutes. Those parameters can be mapped onto values of the axion mass $m$ and decay constant $\mathcal{F}$.

This model has been extensively studied in the literature starting with Refs.~\cite{Poulin:2018cxd,Smith:2019ihp} and has been implemented in a publicly available \href{https://github.com/PoulinV/AxiCLASS}{\faGithub} modified version of the \texttt{CLASS} Boltzmann code \cite{Blas:2011rf}. For more details we refer to the reviews \cite{Kamionkowski:2022pkx,Poulin:2023lkg}.
Early data analyses with Planck 2018 data have shown that the model can provide a good fit to the combination of CMB data and the S$H_0$ES calibrated distance ladder \cite{Poulin:2018cxd}. However, it is not clear that Planck 2018 data independently favor the presence of EDE \cite{Hill:2020osr}, with Bayesian analyses yielding different results than frequentist analyses due to prior-volume effects \cite{Murgia:2020ryi,Smith:2020rxx,Herold:2021ksg}. Interestingly, ACT DR4 data have shown strong preference for the EDE model ($>3\sigma$) over $\Lambda$CDM \cite{Hill:2021yec,Poulin:2021bjr,Smith:2022hwi}, but this appears in tension with results from the latest SPT 3G \cite{LaPosta:2021pgm,Smith:2023oop} and Planck NPIPE data \cite{Efstathiou:2023fbn}.

In this work, we use the results of the latest data analysis of the model in light of Planck NPIPE data (described with the CamSpec likelihood), SDSS BAO and Pantheon$^+$ data calibrated with S$H_0$ES performed in Ref.~\cite{Efstathiou:2023fbn}. For updated analyses that include DESI data, see Refs.~\cite{Qu:2024lpx,Poulin:2024ken}.

\subsection{Scale dependent primordial non-Gaussianities}
\label{sec:spng}
The current best description of the (quantum) generation of the primordial matter fluctuations in accordance with the observed CMB anisotropies requires an inflationary era of at least 60 e-folds of accelerated expansion of our Universe. It predicts an almost scale invariant and almost Gaussian primordial power spectrum in accordance with observations. However, from the model building point of view, UV-complete theories typically predict several fields at low energy generating various features in the inflation power spectrum and in its higher order correlators. These features may be strongly scale-dependent, in relation with specific mechanisms, such as an interaction between the inflaton and a massive scalar field eg.~\cite{Riotto:2010nh, Pinol:2023oux,Goldstein:2024bky}, a tachyonic instability \cite{McCulloch:2024hiz}, how extra-dimensions are wrapped \cite{LoVerde:2007ri} in the context of brane inflation \cite{Chen:2005fe}, vector fields \cite{BeltranAlmeida:2014aau}, or a turn in field space \cite{Khoury:2008wj}. Primordial Black Holes (PBH) formation scenarios are also typical examples where large PNG are present on small scales, in addition to the modifications of the small-scale primordial power spectrum and the isocurvature modes coming from the Poisson fluctuations in the PBH number density \cite{Young:2015kda,Inman:2019wvr,Trashorras:2020mwn,DeLuca:2021hcf,Ferrante:2022mui,Zhang:2024ytf,Coulton:2024vot,Delos:2024poq}.

Contrary to large-scale PNG, that are constrained to be very small, such sPNG are not observationally well constrained, although some CMB \cite{Biagetti:2013sr,Byrnes:2015asa,Oppizzi:2017nfy,Rotti:2022lvy,Sharma:2024img} and LSS \citep[\textit{e.g.,}][]{Dai:2019tjh, Sabti:2020ser, Yamauchi:2021nsf} observables have been used to set loose constraints. Recently, we have also shown \cite{Stahl:2022did, Stahl:2023ccv, Stahl:2024jzk} that a scale dependence in local PNG can impact various galactic-scale and astrophysical quantities. It indeed displays a partial degeneracy with warm/mixed dark matter models at low redshift ($z<3$) \cite{Baldi:2024okt,Stahl:2025nta}. 
Furthermore, sPNG predict a dip in the PS similar to the $A_{\rm mod}$ parametrization of the non-linear suppression \cite{Amon:2022ycy}:
\begin{equation}
\label{eq:Amon}
    P(k,z)=P_{\rm L}(k,z)+A_{\rm mod}\left[P_{\rm NL}(k,z)-P_{\rm L}(k,z) \right],
\end{equation}
where $P(k,z)$ is the matter PS, $P_{\rm L}(k,z)$ and $P_{\rm NL}(k,z)$ are its linear and nonlinear contributions respectively. Instead of directly fitting weak lensing data, or attempting to reproduce the value of the $S_8$ parameter, we will study a sPNG template that reproduces the shape of the PS suppression with $A_{\rm mod}=0.82 \pm 0.04$ found in Ref.~\cite{Preston:2023uup} using KiDS and DES Y3 data.
This approach can be seen as setting a path towards increasing the value of $S_8$ deduced from lensing data, while leaving the linear regime of structure formation untouched \cite{Stahl:2024stz}, which is particularly interesting in the context of EDE models which, on the contrary, primarily affect the background expansion.

Following Ref.~\cite{Stahl:2024stz,Stahl:2025nta}, we will consider the following effective local template for scale-dependent PNG: 
\begin{equation}
\label{eq:fNLk}
f_{\rm NL}(k)=\frac{f_{\rm NL}^0}{1+\alpha}\left[\alpha+\tanh \left(\frac{k-k_{\rm min}}{\sigma} \right) \right],
 \end{equation}
where $\alpha=\tanh \left(\frac{k_{\rm min}}{\sigma} \right)$, and {$f_{\rm NL}^0$, $k_{\rm min}$, $\sigma$} describe the amplitude, scale and shape of the sPNG. Note that our definition for $f_{\rm NL}$ follows from Eq.(4) of Ref.~\cite{Stahl:2024stz}, where $-f_{\rm NL}$ multiplies the quadratic correction to the potential $\Phi$ (and not the curvature $\zeta$; they are related by a factor $3/5$). 

\subsection{Numerical setup}
The initial conditions of our numerical setup are generated using a modified version of the public code \texttt{monofonIC} \href{https://bitbucket.org/ohahn/monofonic/}{\faGithub} \cite{Michaux:2020yis}, customized to include sPNG (Eq.~\eqref{eq:fNLk}) and to be able to call\footnote{To that end, a specific version of \texttt{AxiCLASS} was devised and is available here \href{https://github.com/PoulinV/AxiCLASS/releases/tag/monofonic_v1}{\faGithub} } \texttt{AxiCLASS} for linear cosmology. As sPNG are second order quantities, we used Lagrangian perturbation theory at
second order (2LPT). 

In total, we ran 8 simulations: one benchmark $\Lambda$CDM simulation, one EDE simulation with the model of Sect.~\ref{sec:EDE}, three simulations with the sPNG template of Sect.~\ref{sec:spng}, with $f_{\rm NL}^0 = -300, -600, -1100$. Finally, we ran three joint EDE+sPNG simulations, with the same values of $f_{\rm NL}^0$. In Table \ref{tab:cosmopar}, we display the  cosmological parameters used for each of the simulations studied in this work. In Table \ref{tab:numparam}, we present the numerical choices made for parameters governing the $N$-body simulations. We run these $N$-body simulations until $z=0$ with  \texttt{Gadget-4} \href{https://wwwmpa.mpa-garching.mpg.de/gadget4/}{\faGithub} \cite{Springel:2020plp} and store 5 snapshots at $z=\{3,1,0.5,0.25,0\}$.

\begin{table*}
\begin{center}
\caption{Summary of the simulations performed in this study. The lines give the simulation identifier, the matter density $\Omega_{\rm m}$, baryons density $\Omega_{\rm b}$, the dark energy density $\Omega_{\Lambda}$, the Hubble constant $H_0$ (in km/s/Mpc), the amplitude of the primordial power spectrum $A_s$ and the spectral index $n_s$. They then include four dimensionless parameters for EDE (see Section \ref{sec:EDE}) and three parameters for sPNG: one for its amplitude, $f_{\rm NL}^0$, and two for its shape (Eq.~\ref{eq:fNLk}) $\left\{\sigma,\,k_{\rm min}\right\}$ (in $h$/Mpc).}
\begin{tabular}{|c | c | c | c | c | }
\hline
 & $\Lambda$CDM & EDE & sPNG & EDE+sPNG  \\

\hline 

$\Omega_{\rm m}$ & 0.315 & 0.300 & 0.315 & 0.300  \\
$\Omega_{\rm b}$ & 0.049 & 0.044 & 0.049 & 0.044  \\
$\Omega_{\Lambda}$ & 0.685 & 0.700 & 0.685 & 0.700  \\
$H_0$  & 67.4 & 71.7 & 67.4 & 71.7 \\
$A_s$ & $2.10 \times 10^{-9}$ & $2.14 \times 10^{-9}$ & $2.10 \times 10^{-9}$ & $2.14 \times 10^{-9}$ \\
$n_s$ & 0.967 & 0.988 & 0.967 & 0.988  \\
\hline
$n$ &  & 3 &  & 3 \\
$\log_{10}(a_c)$ &  & -3.566 &  & -3.566 \\
$f_{\rm EDE}$ &  & 0.121 &  & 0.121 \\
$\theta_i$ &  & 2.82 &  & 2.82 \\
\hline
$f_{\rm NL}^0$ &  &  & -300/-600/-1100 & -300/-600/-1100 \\
$k_{\rm min}$ &  &  & 0.15 & 0.15  \\
$\sigma$ &  &  & 0.1 & 0.1  \\

\hline
\end{tabular}
\label{tab:cosmopar}
\vspace{-5mm}
\end{center}
\end{table*}

\begin{table*}
\begin{center}

\caption{Description of the numerical parameters for all the simulations presented in this work: box length $L_{\rm box}$ (in Mpc/$h$), number of particles $N_{\rm part}$, starting redshift $z_{\rm start}$, particles' mass $m_{\rm part}$ (in $M_{\odot}$/$h$), softening length $L_{\rm soft}$ (in kpc/$h$).}
\begin{tabular}{ |c| c | c | c |  c |}
\hline  $L_{\rm box}$  & $N_{\rm part}$  & $z_{\rm start}$ & $m_{\rm part}$ &  $L_{\rm soft}$ \\ 

\hline 

 500 & $512^3$ & 32 & $8 \times 10^{10}$ &  50 \\

\hline
\end{tabular}
\label{tab:numparam}
\vspace{-5mm}
\end{center}
\end{table*}

\section{Results}\label{sec:res}

We now present the results of the simulations described in Table \ref{tab:cosmopar}. We study two key quantities of structure formation: the matter power spectrum (PS) and the Halo Mass Function (HMF). Simulations of EDE were already carried out in Refs.~\cite{Klypin:2020tud,Murgia:2020ryi} while simulations of sPNG were presented in Refs.~\cite{Stahl:2024stz,Baldi:2024okt, Stahl:2025nta}. Our results for EDE and sPNG simulated separately are in agreement with those previous works.

\subsection{Matter power spectrum}
\label{sec:PS}

\begin{figure}

        \includegraphics[width=\textwidth]{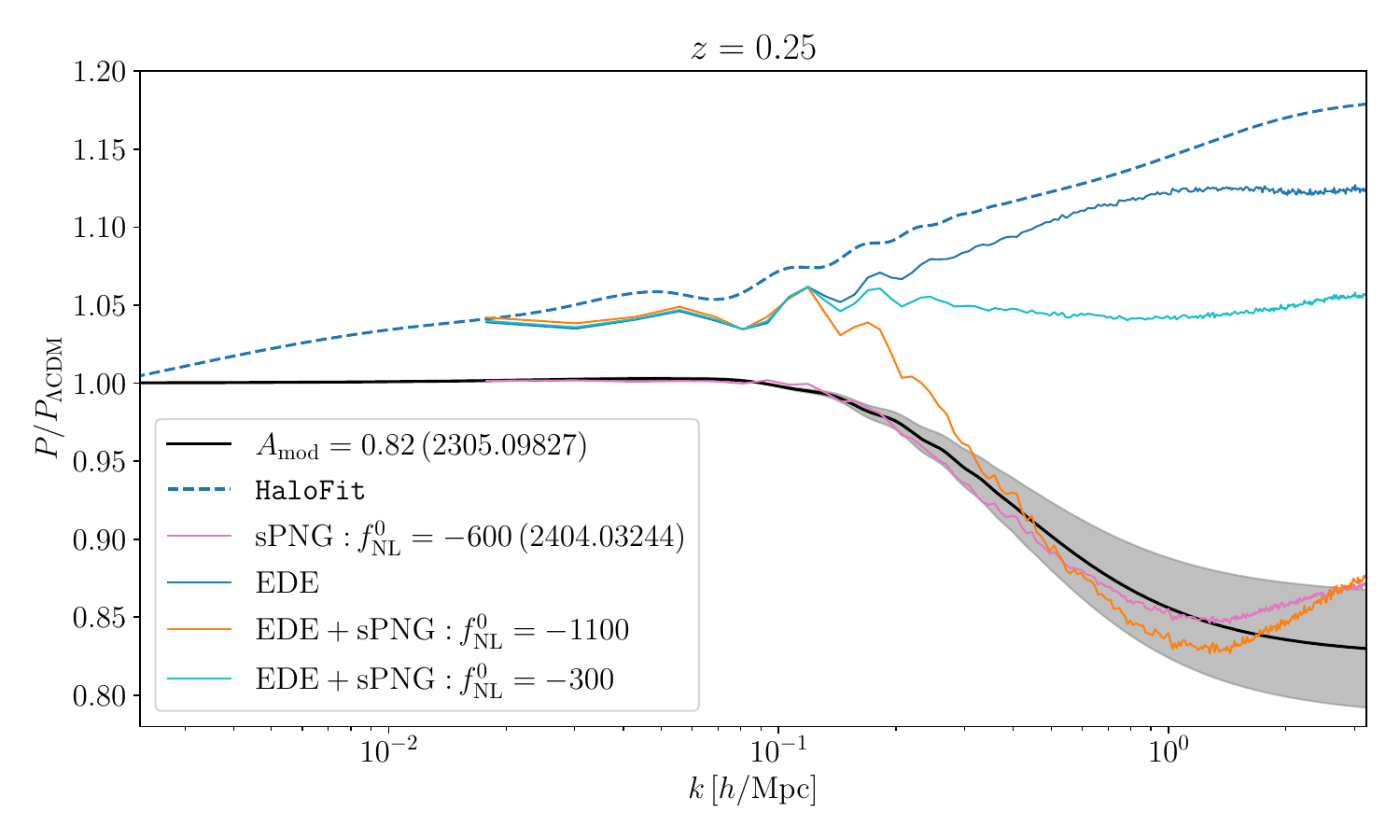}
                \caption{Ratio of the PS for different simulations studied in this work (cf.~Table \ref{tab:cosmopar}) at redshift $z=0.25$.}
                      \label{fig:PS}
         \end{figure}

In Fig.~\ref{fig:PS}, we present the ratio of the PS of the simulations studied in this work to that of $\Lambda$CDM at $z=0.25$. We contrast it to the non-linear solution to the $S_8$ tension exhibited by Ref.~\cite{Preston:2023uup}. Redshift 0.25 is where the statistical power of the DES Y3 + KiDS surveys is maximal.

The wave-number range which is relevant to $S_8$ measurements spans from $k \sim 0.2 \, h/{\rm Mpc}$ to $k \gtrsim 1 \, h/{\rm Mpc}$, while for $k<0.2 \, h/{\rm Mpc}$, observational constraints on the PS are of the order of 5\% and all the models considered agree with CMB constraints. Figure~\ref{fig:PS} displays a sPNG model already known to solve the original $S_8$ tension \cite{Stahl:2024stz}, with $f_{\rm NL}^0=-600$, together with our present EDE model (see Sect.~\ref{sec:EDE}) which clearly worsens the $S_8$ tension, and a combination of EDE and sPNG with $f_{\rm NL}^0=-300$, which flattens the PS at high $k$ compared to EDE alone, and with $f_{\rm NL}^0=-1100$, which allows to recover a PS at high $k$ compatible with what is needed to resolve the original $S_8$ tension. In the EDE-alone case, we also provide the theoretical PS from the \texttt{Halofit} emulator, which overestimates the PS compared to the simulation, due to the finiteness of the simulated scales (cosmic variance, see also \cite{Murgia:2020ryi}).

This thus demonstrates that combining sPNG with EDE models designed to solve the Hubble tension can, at the very least, bring back the EDE models in line with $\Lambda$CDM regarding the $S_8$ parameter, or even solve both the Hubble and the original $S_8$ tensions together with a somewhat `aggressive' value of $f_{\rm NL}^0$. Furthermore, this (re-)demonstrates that the constraints derived on EDE from LSS data are not robust to change to the small-scale matter power spectrum (see also Refs.~\cite{Clark:2021hlo,Simon:2024jmu}), in our case motivated by a change to the initial conditions of the matter density fluctuations.

\begin{figure}

        \includegraphics[width=\textwidth]{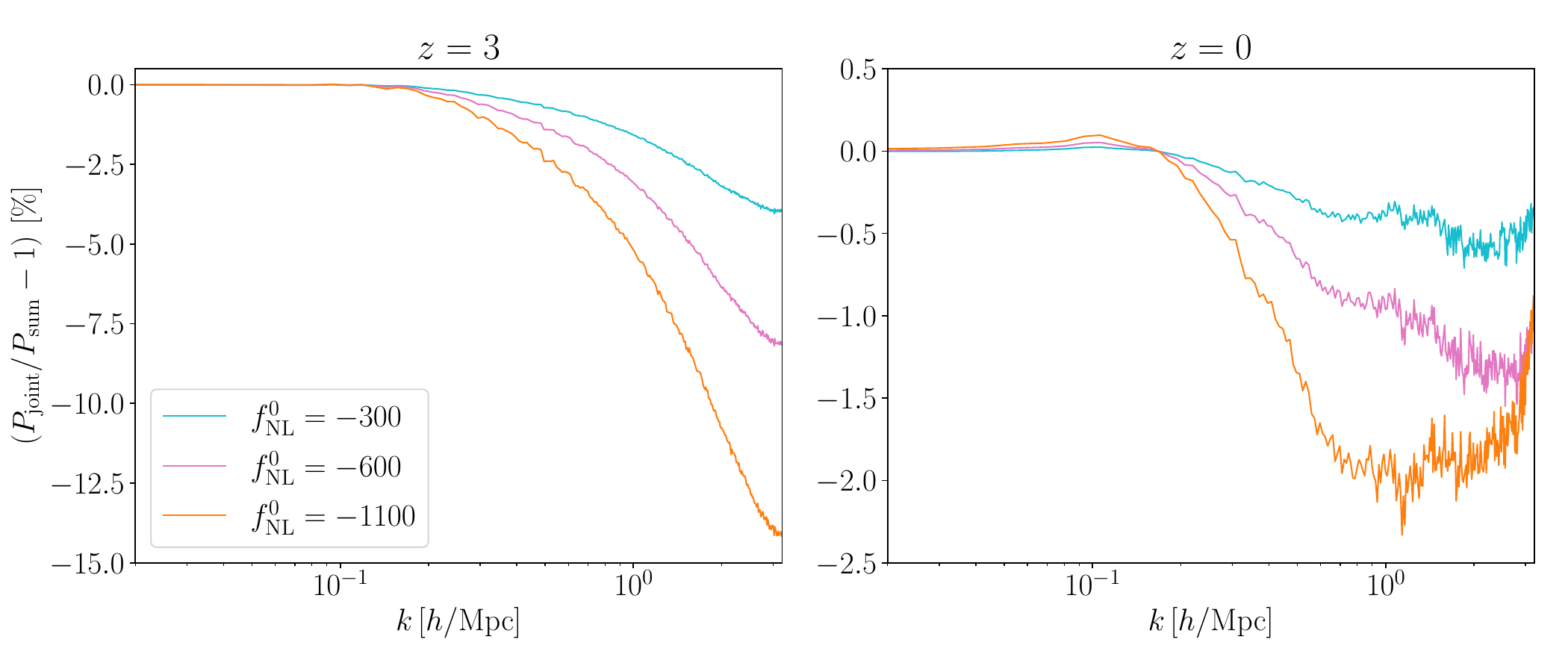}
                \caption{Ratio of the joint PS (EDE and sPNG simulated together), $P_{\rm joint}$, with respect to the linear combination $P_{\rm sum}$ described in Eq.~\eqref{eq:Psum}. At $z=3$, $P_{\rm sum}$ is an excellent approximation at the largest scales, whilst at $z=0$ it approximates $P_{\rm joint}$ with percent accuracy at all scales.}
                      \label{fig:jVSs}
         \end{figure}

These results are qualitative, as firmly establishing the region of parameter space required to solve cosmic tensions is beyond the scope of this paper and requires a dedicated emulator trained on simulations with non-trivial primordial PS. For this, it is most useful to have a quick way to estimate the joint effect of EDE and sPNG on the PS. Simply summing up the two effects would greatly simplify the task of building an emulator for the non-linear corrections due to sPNG, as the EDE parameters would not need to be included as free parameters in the latin hypercube of the emulator. We therefore check hereafter whether the sum of the PS compared to the $\Lambda$CDM case can appropriately reproduce the joint effect of EDE and sPNG on the PS: 
\begin{equation}
         \label{eq:Psum}
             P_{\rm sum} \equiv P_{\rm EDE}+P_{\rm sPNG}-P_{\rm \Lambda CDM}.
         \end{equation}
In Fig.~\ref{fig:jVSs}, we display the ratio of the PS of the joint EDE \& sPNG simulations over the linearly superposed PS $P_{\rm sum}$. In the left panel, at $z=3$, the linear superposition underestimates the joint effect of sPNG and EDE at the smallest scales simulated, by up to 15\% at $k>1 \, h/{\rm Mpc}$ for the largest value of $f_{\rm NL}^0$. The discrepancy is significantly smaller on large scales and, interestingly, becomes also much smaller at lower redshifts, for instance at the percent level at $z=0$, as shown on the right panel of Fig.~\ref{fig:jVSs}.

\subsection{Halo Mass Function}
\label{sec:HMF}

\begin{figure}

        \includegraphics[width=\textwidth]{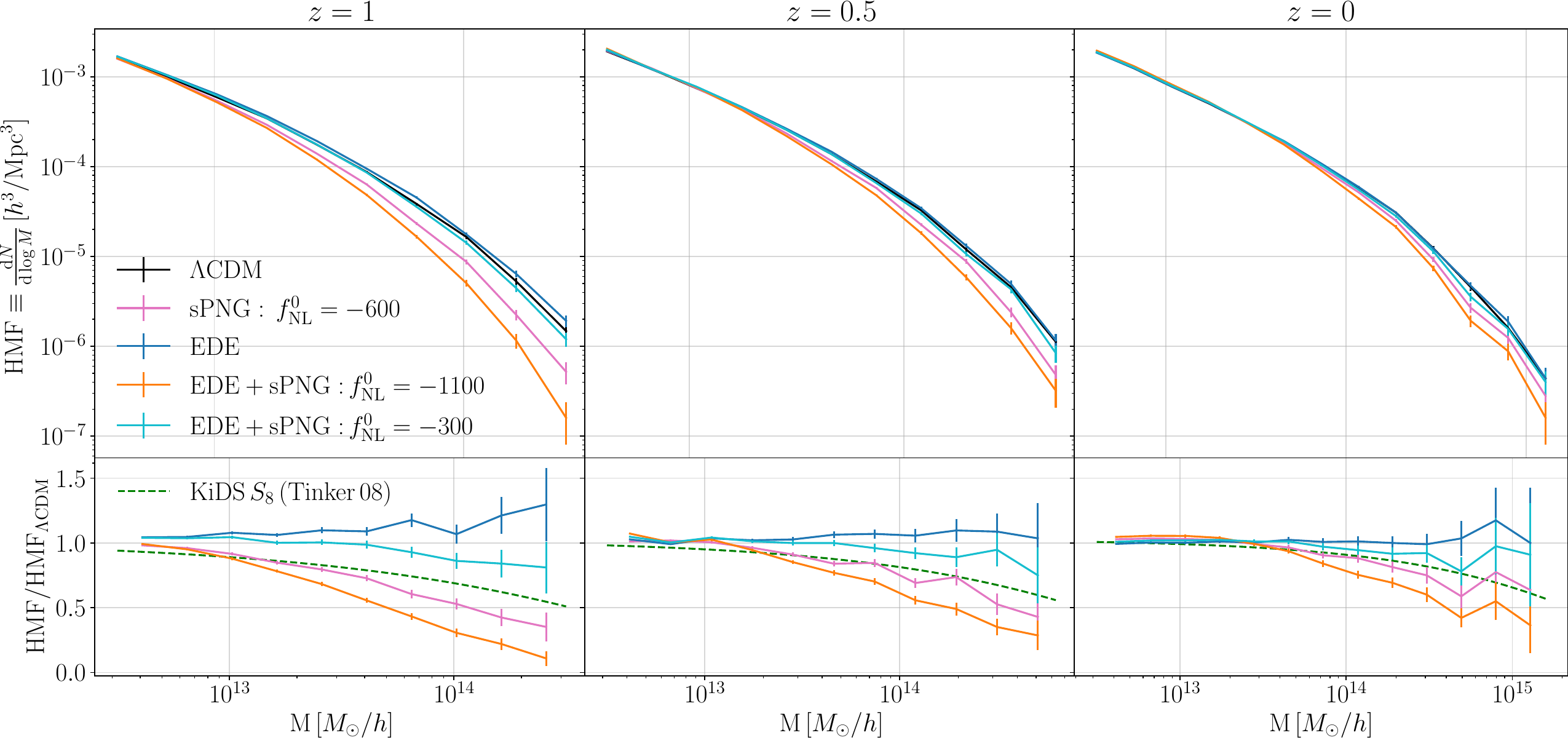}
                \caption{Top row: HMF of the $\Lambda$CDM, EDE, sPNG, and joint EDE \& sPNG simulations at $z=1, 0.5, 0$. Bottom row: Ratio of the HMF with respect to the Planck $\Lambda$CDM one. The HMF emulated with the fitting formula of Ref.~\cite{Tinker:2008ff} for a $\Lambda$CDM Universe with the $S_8$ value inferred from KiDs is also overplotted. Error bars are Poissonian.}
                     \label{fig:HMF}
         \end{figure}

Halos were identified using the \texttt{Subfind} tool included in the public release of \texttt{Gadget-4}. Each halos’s boundary is determined where its density reaches 200 times the critical density. Halos with fewer than 100 particles were discarded from the analysis. The HMF of the different simulations are presented on Fig.~\ref{fig:HMF}. As previously noted the sPNG simulation with $f_{\rm NL}^0 =-600$ not only brings the PS in accordance to what is needed to solve the original $S_8$ tension but also solves it in the context of cluster counts \cite{Planck:2015lwi}. Conversely, the EDE-alone simulation boosts the HMF \citep[see also][]{Klypin:2020tud,Shen:2024hpx}: the higher the redshift, the higher the boost of the HMF. The joint EDE \& sPNG simulation with a large value of $f_{\rm NL}^0=-1100$ leads to a stronger suppression of the HMF for high masses. We leave for future work the dedicated analyses to constrain the effect of sPNG on observed cluster catalogs. The joint EDE \& sPNG simulation $f_{\rm NL}^0=-300$ that flattens the EDE PS also stays very close to $\Lambda$CDM in terms of the HMF at $z \leq 1$. Finally, such a simulation is expected to boost the PS at high redshifts ($z > 8$): although we do not have enough resolution to isolate halos at such early epochs, we expect that it would help boost the formation of halos at such redshifts \cite{Klypin:2020tud}, which might be interesting in the context of the JWST observations of the formation of massive galaxies at high redshift \cite{Boylan-Kolchin:2022kae,Forconi:2023hsj,Shen:2024hpx}. 

Concerning cluster counts, it is worthwhile to note that these were recently updated \cite{Ghirardini:2024yni} by the eROSITA collaboration (with clusters typically in the range from $2 \times 10^{13} \, {\rm M}_\odot$ to a few $10^{14} \, {\rm M}_\odot$), which provided a value of $S_8$ {\it higher} than the $\Lambda$CDM value, $S_8 = 0.86\pm0.01$, in significant tension with the old weak-lensing estimates and the previous cluster count constraints. Understanding the origin of this difference is beyond the scope of the present work, but we note that our joint EDE \& sPNG simulation with a large value of $f_{\rm NL}^0=-1100$ would also be in tension with this result, as it leads to significantly less clusters than in $\Lambda$CDM. The simulation with EDE and $f_{\rm NL}^0=-300$, which would bring back the PS and HMF mostly in accordance with $\Lambda$CDM, would on the other hand be closer to these eROSITA results, as well as to the latest KiDS results \citep{Wright:2025xka,Stolzner:2025htz}, while still solving the $H_0$ tension.

Finally, let us note that the linear superposition of Eq.\eqref{eq:Psum} for the PS also works well (at the percent level at $z=0$) for the HMF.

\section{Conclusions and perspectives}\label{sec:ccl}

In this paper, we demonstrated the potential of combining EDE with sPNG to reconcile the $H_0$ and $S_8$ parameters. We have shown that a joint $N$-body simulation of sPNG and EDE suppresses the non-linear PS relevant to the $S_8$ parameter estimate from weak-lensing, while preserving the EDE solution to the Hubble tension. Depending on the amplitude of the sPNG, the $S_8$ parameter of EDE models can be brought back in line with $\Lambda$CDM expectations, while still addressing the Hubble tension. The analysis of the HMF further supports these findings. However, the joint EDE \& sPNG simulation solving the original $S_8$ tension inferred from weak-lensing would likely not explain the recent results of the eROSITA collaboration \cite{Ghirardini:2024yni}. The  EDE \& sPNG simulation bringing back the non-linear PS at the $\Lambda$CDM level would on the other hand be closer to these eROSITA results, as well as to the latest results from KiDS \citep{Wright:2025xka,Stolzner:2025htz}. Broadening the perspective, it was also suggested that the boost in the HMF at high-redshift in the EDE cosmology can help explain the JWST results on the formation of massive galaxies at high-redshift \cite{Boylan-Kolchin:2022kae,Forconi:2023hsj,Shen:2024hpx}. Since our simulations show that sPNG significantly affect the PS only at low redshift, it is likely similar for the HMF: hence the combination of sPNG and EDE studied in this work could potentially also explain JWST observations of high-redshift massive galaxies. The design and analysis of much higher resolution simulations that could resolve halos at high redshifts should be the topic of subsequent works. In that context, sPNG may also come along with primordial black holes, that could also ease the formation of high-$z$ galaxies \cite{Colazo:2024jmz, Huang:2024aog}.

Our results are qualitative, and firmly establishing the region of parameter space that solve the tensions would require a dedicated emulator trained on simulations with non-trivial primordial physics. In preparation to construct such a tool, we demonstrated that the sum of the separated EDE and sPNG power spectra compared to the $\Lambda$CDM case can appropriately reproduce the joint effect of EDE and sPNG on the PS, at the percent level at low redshifts. Before emulating the simulated physics, we will need to refine the theoretical templates, as inflationary models typically predict features affecting only a localized range of scales. They would boost or damp the initial matter fluctuations both at the level of the primordial PS and bispectrum. An example derived from UV-complete models of inflation was presented in Ref.~\cite{Stahl:2025qru}, and the effect on the primordial PS was shown to be similar to a typical drop induced by warm/mixed dark matter candidates. Such effects may contribute jointly with sPNG in solving cosmic tensions (along, or not, with baryonic effects). The construction of a powerful emulator will require to have all these ingredients neatly combined, using linear superpositions when possible.

In summary, we have shown that the freedom in the form of the primordial matter distribution, ultimately connected with our currently limited constraints on the physics of the inflation era,
is an important caveat to the robustness of LSS-driven constraints to EDE. We have indeed presented a compelling case for considering the combined effects of EDE and sPNG in resolving cosmological tensions, offering a pathway towards a more comprehensive understanding of the primordial Universe.

\acknowledgments
We thank Oliver Hahn for initial help to plug \texttt{AxiCLASS} to \texttt{monofonIC}. VP is supported by funding from the European Research Council (ERC) under the European Union’s HORIZON-ERC-2022 (grant agreement No 101076865).
This work has made use of the Infinity Cluster hosted by the Institut d'Astrophysique de Paris.

\section*{Softwares}
The analysis was partially made using \texttt{YT} \href{https://yt-project.org/}{\faGithub} \cite{Turk:2010ah}, \texttt{Pylians} \href{https://pylians3.readthedocs.io/en/master/index.html}{\faGithub} \cite{Pylians}, \texttt{Colossus} \href{https://bdiemer.bitbucket.io/colossus/index.html}{\faGithub} \cite{Diemer:2017bwl} as well as IPython \cite{Perez:2007emg}, Matplotlib \cite{Hunter:2007ouj} and NumPy \cite{vanderWalt:2011bqk}.

\section*{Authors' Contribution}
VP and CS designed the initial conditions for the simulations presented in this work that were performed and analyzed by CS in consultation with BF. CS and BF drafted the manuscript with inputs from VP. All the authors reviewed, discussed, and provided feedback on the manuscript.

\section*{Carbon Footprint}
In Ref.~\cite{berthoud}, 1 hour core has been shown to be equivalent to 3.6 gCO2e including the global usage of a cluster and the pollution due to the electrical source. The simulations for this work required then 0.3 TCO2eq.

\bibliographystyle{JHEP.bst}
\bibliography{ref.bib}

\end{document}